# Additive Layers: An Alternate Classification Of Flow Regimes

Trinh, Khanh Tuoc

Institute of Food Nutrition and Human Health

Massey University, New Zealand

K.T.Trinh@massey.ac.nz

## Abstract

It is argued that ejections of wall fluid in the bursting process disturb the flow beyond the wall layer and result in the emergence of two new layers in the flow field: the law of the wake and log-law layers. The wall layer represents the extent of penetration of viscous momentum into the main flow and remains constant once it has reached a critical value at the end of the laminar regime. The identification of the three flow regimes: laminar, transition and fully turbulent is conveniently achieved by monitoring the emergence of these three layers

## 1 Introduction

The scientific literature distinguishes two distinct flow regimes laminar and turbulent. In a classic visual experiment Reynolds (1883) observed that dye streaks introduced into the flow followed clear streamlines in laminar flow but became diffuse at high fluid velocity. He proposed that turbulence set in when the Reynolds number reaches a critical value

$$\text{Re} = \frac{DV\rho}{\mu} \tag{1}$$

where $D$ is the pipe diameter, $V$ the average velocity, $\rho$ the fluid density and $\mu$ the dynamic viscosity. For pipe flow, he showed that the laminar flow regime ends at $\text{Re}_c = 2100$. The characteristic dimension and velocity change with the system geometry and so does the critical Reynolds number. For flow past a sphere

$$\text{Re}_c = \frac{D_p U_\infty \rho}{\mu} = 1 \tag{2}$$

Reynolds (1895) has proposed that the instantaneous velocity $u_i$ at any point may be decomposed into a long-time average value $U_i$ and a fluctuating term $U'_i$.

$$u_i = U_i + U'_i \tag{3}$$

with

$$U_i = \lim_{t \to \infty} \int_0^t u_i \, dt \tag{4}$$

$$\int_0^\infty U'_i \, dt = 0 \tag{5}$$

For simplicity, we will consider the case when

1. The pressure gradient and the body forces can be neglected

2. The fluid is incompressible ($\rho$ is constant). Substituting equation (3) into the Navier-Stokes equations

$$\frac{\partial}{\partial t}(\rho u_i) = -\frac{\partial}{\partial x_i}(\rho u_i u_j) - \frac{\partial}{\partial x_i} p - \frac{\partial}{\partial x_i} \tau_{ij} + \rho g_i \tag{6}$$

and taking account of the continuity equation

$$\frac{\partial}{\partial x_i}(\rho u_i) = 0 \tag{7}$$

gives:

$$U_i \frac{\partial U_j}{\partial x_j} = \nu \frac{\partial^2 U_i}{\partial x_i} - \frac{\partial \overline{U'_i U'_j}}{\partial x_i} \tag{8}$$

These are the famous Reynolds-Averaged Navier-Stokes equations (Schlichting 1960, p. 529). The long-time-averaged products $\overline{U'_i U'_j}$ arise from the non-linearity of the Navier-Stokes equations. They have the dimensions of stress and are known as the Reynolds stresses. They are absent in steady laminar flow and form the distinguishing features of turbulence. The Reynolds number is usually interpreted as a ratio of the turbulent and viscous driving forces in the flow. It is not often mentioned that Reynolds also observed that the diffuse mass of dye in turbulent flow seen "by the light of an electric spark" is actually organised as "a mass of more or less distinct curls, showing eddies" as shown in Figure 1. Thus Reynolds was the first to detect what are now called coherent structures in turbulent flow fields.

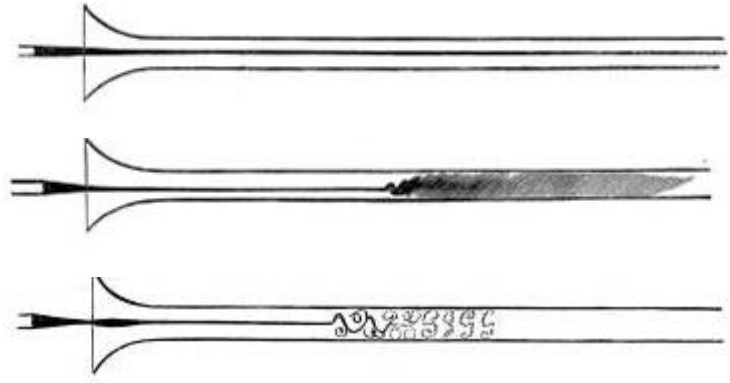

Figure 1. Dye traces in laminar and turbulent flows. From Reynolds (1883). Top: laminar flow; middle: diffuse dye viewed by the naked eye; bottom: coherent structures seen under electric spark

## 2  Basic considerations

Eighty years later Kline, Reynolds, Schraub, & Runstadler (1967) reported their now classic hydrogen bubble visualisation of events near the wall and renewed interest in the coherent structures. Despite the prevalence of viscous diffusion of momentum close to the wall, the flow was not laminar in the steady state sense envisaged by Prandtl (1935). The low-speed streaks tended to lift, oscillate and eventually eject away from the wall in a violent burst. In side view, they recorded periodic inrushes of fast fluid from the outer region towards the wall followed by a vortical sweep along the wall. The low-speed streaks appeared to be made up of fluid underneath the travelling vortex as shown in Figure 2. The bursts can be compared to jets of fluids that penetrate into the main flow, and get slowly deflected until they become eventually aligned with the direction of the main flow. The observations of Kline et al. have been confirmed by many others e.g. (Corino & Brodkey, 1969; H. T. Kim, Kline, & Reynolds, 1971; Offen & Kline, 1974, 1975).

Because of the importance of the wall region as highlighted by the work of Kline et al., a large amount of effort has been devoted to its study focussing mainly on the hairpin vortex, the most identifiable coherent structure in that region. Work before

1990 were well reviewed, for example by Cantwell (1981) and Robinson (1991). There have been physical experiments e.g. (Blackwelder & Kaplan, 1976; Bogard & Tiederman, 1986; Carlier & Stanislas, 2005; Corino & Brodkey, 1969; Head & Bandhyopadhyay, 1981; Luchak & Tiederman, 1987; Meinhart & Adrian, 1995; Tardu, 2002; Townsend, 1979; Willmarth & Lu, 1972), including efforts to induce artificially the creation of a hairpin vortex by injecting a jet of low momentum fluid into a laminar flow field (Arcalar & Smith, 1987; Gad-el-Hak & Hussain, 1986; Haidari & Smith, 1994). With the advent of better computing facilities, direct numerical simulations DNS have been used increasingly to conduct 'numerical experiments" e.g. (Jimenez & Pinelli, 1999; J. Kim, Moin, & Moser, 1987; Spalart, 1988).

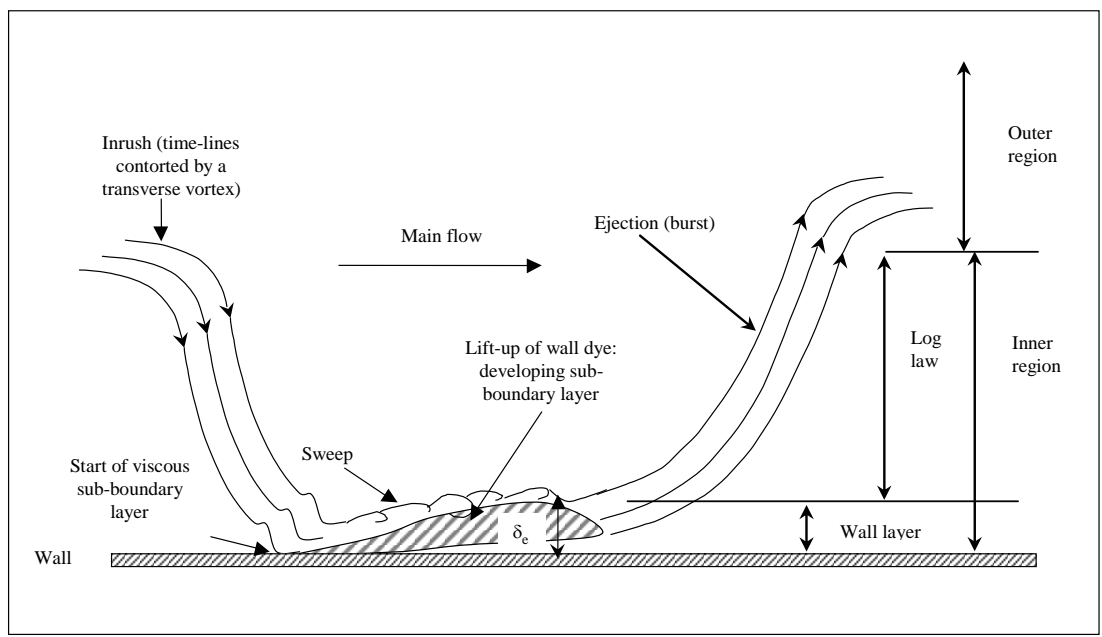

Figure 2. The wall layer process drawn after Kline et al. (1967) and Kim et al. (1991).

There is wide consensus that the presence of a longitudinal vortex above the wall and a bursting process are necessary for the onset of turbulent flow. Kernel studies that simulate the passage of a vortex above a wall as a model for the turbulent wall process e.g. (Peridier, Smith, & Walker, 1991; Swearingen & Blackwelder, 1987; Walker, 1978) showed that the vortex induces an oscillating sub-boundary layer under its path (sweep phase) that erupts in a violent so called viscid-inviscid interaction (burst, ejection) when the fluctuations have grown sufficiently. The ejection reaches well beyond the sub-boundary layer induced. The sweep phase can be modelled as a

Kolmogorov flow (Obukhov, 1983) a simple two dimensional sinusoidal flow, or better still analysed with techniques borrowed from laminar oscillating flow (K.T. Trinh, 1992; K. T. Trinh, 2009). The velocity in the sweep phase can be decomposed into a smoothed phase velocity $\tilde{u}_i$ and fast fluctuating component $u'_i$. The traditional approach to analyse these unsteady flows is by a method of successive approximations (Schlichting, 1979; Tetlionis, 1981). The dimensionless parameter defining these successive approximations is

$$\varepsilon = \frac{U_e}{L\omega} \tag{9}$$

where $U_e$ is the local mainstream velocity and L is a characteristic dimension of the body. The smoothed velocity $\tilde{u}_i$ is given by the solution of order $\varepsilon^0$ which applies when $\varepsilon \ll 1$. The governing equation (Einstein & Li, 1956; Hanratty, 1956; Meek & Baer, 1970; K. T. Trinh, 2009) is a subset of the NS equations

$$\frac{\partial \tilde{u}}{\partial t} = \nu \frac{\partial^2 \tilde{u}}{\partial y^2} \tag{10}$$

where $\tilde{u}$ refers here to the smoothed velocity $\tilde{u}_i$ in the $x$ direction. It does not require that there are no velocity fluctuations, only that they are small enough for their effect on the smoothed phase velocity $\tilde{u}$ to be negligible. Stokes (1851) has given the solution to equation (10) as

$$\frac{\tilde{u}}{U_\nu} = \text{erf}(\eta_s) \tag{11}$$

where $\eta_s = \frac{y}{\sqrt{4\nu t}}$. The thickness of this sub-boundary layer is

$$\delta_\nu^+ = 4.16 U_\nu^+ \tag{12}$$

where the velocity and normal distance have been normalised with the wall parameters $\nu$ the kinematic viscosity and $u_* = \tau_w/\rho$ the friction velocity, $\tau_w$ the time averaged wall shear stress and $\rho$ the density.

As the Reynolds number increases, the magnitude of the velocity fluctuations grows ($\varepsilon$ increases) according to well-known analyses of stability of laminar flows e.g. (Dryden, 1934, 1936; Schiller, 1922; Schlichting, 1932, 1933, 1935; Schubauer & Skramstad, 1943; Tollmien, 1929). We then switch to a second approximation of order $\varepsilon$. We may average the Navier-Stokes equations over the period $t_f$ of the fast fluctuations. Bird, Stewart and Lightfoot (1960, p. 158) give the results as

$$\frac{\partial(\rho \tilde{u}_i)}{\partial t} = -\frac{\partial p}{\partial x_i} + \mu \frac{\partial^2 \tilde{u}_i}{\partial x_j^2} - \frac{\partial \tilde{u}_i \tilde{u}_j}{\partial x_j} - \frac{\partial \overline{u'_i u'_j}}{\partial x_j} \tag{13}$$

Equation (13) defines a second set of Reynolds stresses $\overline{u'_i u'_j}$ which we will call "fast" Reynolds stresses to differentiate them from the standard Reynolds stresses $\overline{U'_i U'_j}$.

Within a period $t_v$, the smoothed velocity $\tilde{u}_i$ varies slowly with time but the fluctuations $u'_i$ may be assumed to be periodic with a timescale $t_f$. In the particular case of steady laminar flow, $\tilde{u}_i = U_i$ and $\tilde{U}'_i = 0$: only the fast fluctuations $u'_i$ remain. These are typically remnants of disturbances introduced at the pipe entrance or leading edge of a flat plate by conditions upstream.

We may write the fast fluctuations in the form

$$u'_i = u_{o,i}(e^{i\omega t} + e^{-i\omega t}) \tag{14}$$

The fast Reynolds stress $u'_i u'_j$ becomes

$$u'_i u'_j = u_{o,i} u_{o,j}(e^{2i\omega t} + e^{-2i\omega t}) + 2u_{o,i} u_{o,j} \tag{15}$$

Equation (15) shows that the fluctuating periodic motion $u'_i$ generates two components of the "fast" Reynolds stresses: one is oscillating and cancels out upon long-time-averaging, the other, $u_{0,i} u_{0,j}$, is persistent in the sense that it does not depend on the period $t_f$. The term $u_{0,i} u_{0,j}$ indicates the startling possibility that a purely oscillating motion can generate a steady motion which is not aligned in the direction of the oscillations. The qualification steady must be understood as independent of the frequency $\omega$ of the fast fluctuations. If the flow is averaged over a longer time than the period $t_v$ of the bursting process, the term $u_{0,i} u_{0,j}$ must be understood as transient but non-oscillating. This term indicates the presence of transient shear layers embedded in turbulent flow fields and not aligned in the stream wise direction similar to those associated with the streaming flow in oscillating laminar boundary layers (Schneck & Walburn, 1976; Tetlionis, 1981). Schneck and

Walburn (1976) have argued in their study of pulsatile blood flow that the secondary streaming flow results from a tendency of viscous forces to resist the reversal of flow imposed by the oscillating motion of the main stream. This is demonstrated more clearly in the experiments of Gad-el-Hak, Davis, McMurray, & Orszag (1983) who generated an artificial bursting process in a laminar boundary layer on a flat plate by decelerating it. The magnitude of the deceleration and the corresponding adverse pressure gradient must be sufficient to induce separation and ejection of low-speed fluid from the wall.

At the end of the sweep phase, the fluctuations have grown large enough for the streaming flow to contain substantial amount of kinetic energy sufficient to eject wall fluid from the wall layer. The ejections start to disturb the outer quasi-inviscid region beyond the wall layer and dramatically increase the boundary layer thickness. At Reynolds numbers just above the critical value, e.g. Re =2100 for pipe flow, only the far field section of the intermittent jets penetrates the outer region. The disturbance to the previously "quasi-potential" flow may be compared with that of a wall-parallel jet since the ejections are here aligned in the direction of main flow. This region has been described by Cole's law of the wake (Coles, 1956). As the Reynolds number increases further, so do the fluctuations: the streaming flow strengthens and emerges at a cross flow angle with the main flow. Millikan (1939) showed that an outer region that scale with the outer parameters (Coles law of the wake in the present visualisation) and a wall region which scales with the wall parameters (the solution of order $\varepsilon^0$) must be linked by a semi-logarithmic velocity profile.

$$U^+ = A \ln y^+ + B \qquad (16)$$

In the present visualisation, upon transition, the first layer to be added to the wall layer is the law-of-the-wake region then full turbulence is established when the log-law grows.

The transition from laminar to turbulent flow is clearly seen in the oscillating flow experiments of Akhavan, Kamm, & Saphiro Akhavan (1991) . The flow is driven by a reciprocating piston pump. The acceleration phase, where the pressure gradient is favourable, is laminar. The velocity profile here exhibits only two regions: (a) a wall layer which coincides very well with the profiles for laminar boundary layer flow and those for the wall layer of steady turbulent pipe flow, and (b) a fluctuating potential flow in the outer region.

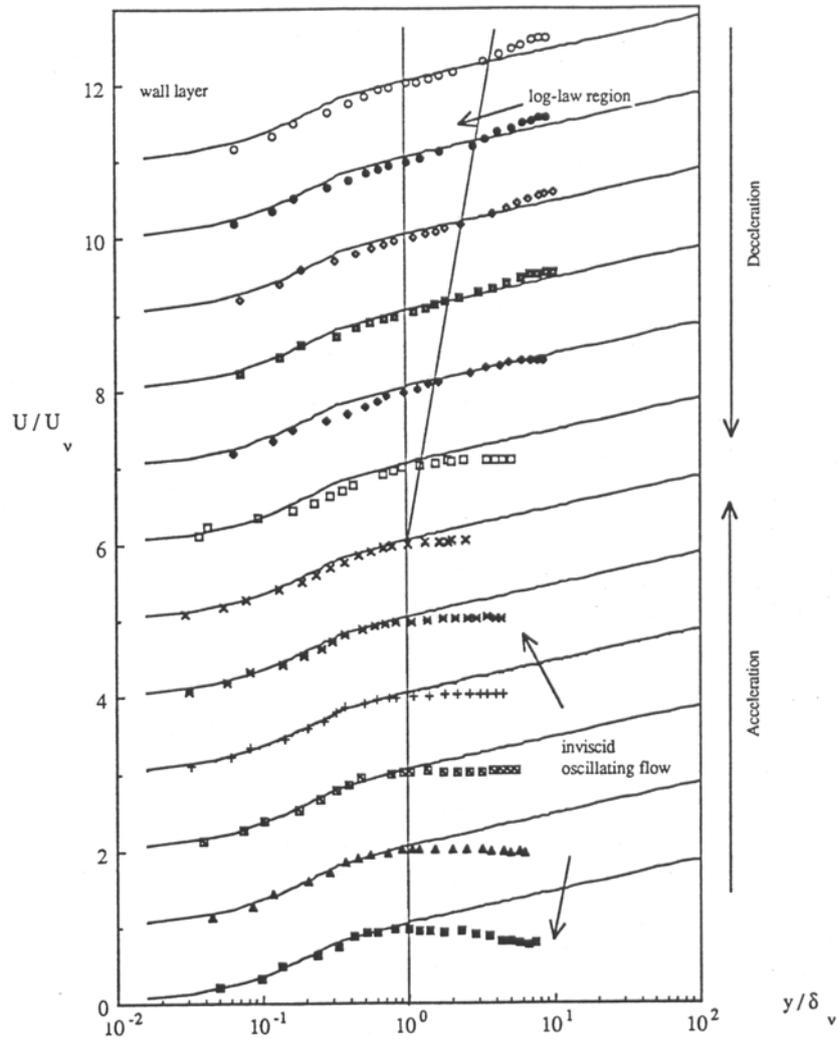

Figure 3. Penetration of the log-law into the outer region during a cycle oscillating pipe flow. From Trinh (1992). Data of Akhavan et al. (1991) for $Re_\omega = 1080$.

The growth of the log-law region in between the wall layer and the law of the wake during the decelerating phase of oscillating flow is seen clearly in Figure 3 where the original data of Akhavan et al. has been rearranged (K.T. Trinh, 1992; K. T. Trinh, 2009).

## 3 Determination of layer boundaries

The boundary of different layers was determined by plotting the measured velocity profile in semi-log plot as shown in Figure 4.

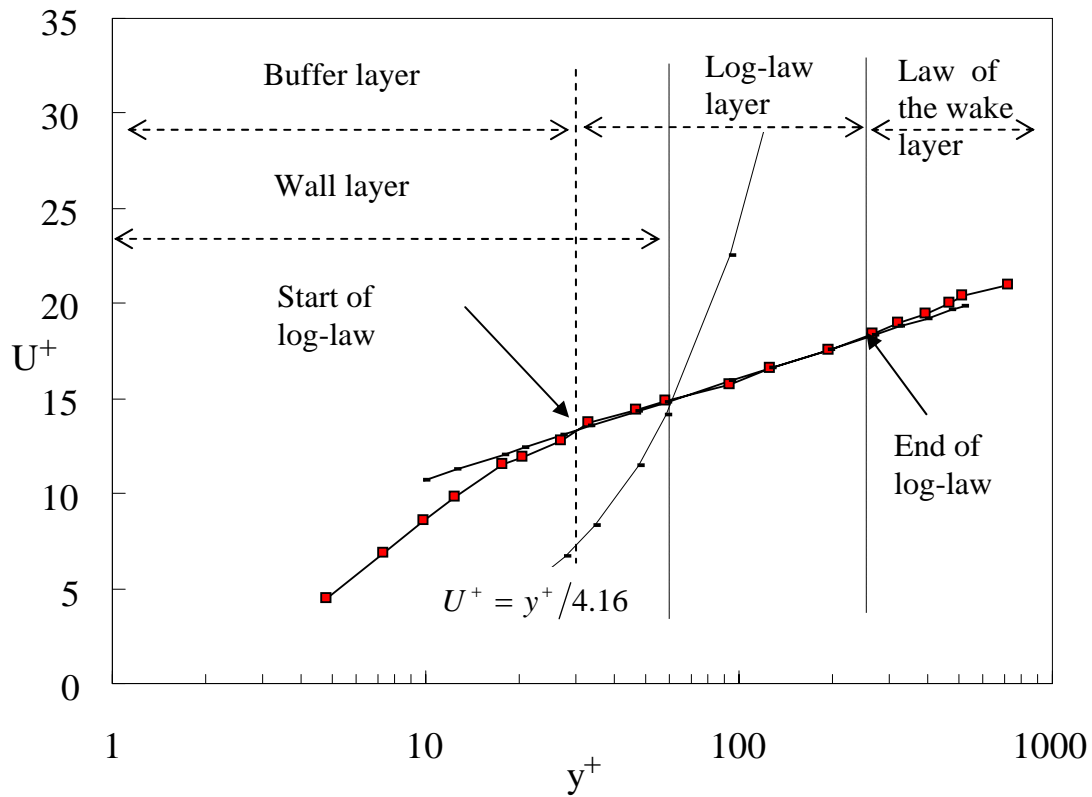

Figure 4. Determination of layers boundaries. Data of Wei and Wilmarth (1989).

A straight line was fitted to the mid portion. The point where it first coincides with the experimental data is taken as the lower boundary of the log-law. Two points can be identified: the edge of the wall layer which represents the maximum penetration of viscous momentum from the wall into the main flow and the edge of the buffer layer which represents the time-averaged penetration of viscous momentum (Trinh, 2009, p.52). The point where the log-law again departs from the experimental measurements is the boundary between the log-law and law of the wake layers. Figure 5 has been compiled from velocity measurements of boundary layer flow by Klebanoff (1954) and others (Schlichting, 1960) and pipe flow by various authors (Bogue, 1962; Eckelmann, 1974; Laufer, 1954; Lawn, 1971; Nikuradse, 1932; Senecal & Rothfus, 1953). The top graph (Figure 5 a) presents the four traditional regimes of flow past a flat plate. Figure 5 (b) shows typical velocity traces for the four regimes from

Schubauer and Skramstadt (op.cit.). Figure 5 (c) shows the normalised thicknesses of the three layers: Karman buffer layer (or time averaged wall layer) log-law and law of the wake. Figure 5 (d) shows the physical thicknesses of the three layers (not to scale) and the shape of the streaming flow at different Reynolds numbers.

The edge of the boundary layer is given by the last point on the velocity curve. It can also be calculated in terms of the Reynolds number $\text{Re}_x$ from the friction factor equation

$$f = \frac{\alpha}{\text{Re}_\delta^\beta} \tag{17}$$

and the velocity profile

$$\frac{U}{U_\infty} = \left(\frac{y}{\delta}\right)^p \tag{18}$$

The indices $p$ and $\beta$ are related (Schlichting 1960)

$$p = \frac{\beta}{2-\beta} \tag{19}$$

A derivation by standard methods (Schlichting p.598) gives the boundary layer thickness as

$$\frac{\delta}{x} = \left[\frac{2\alpha(\beta+1)(\beta+1)}{2-\beta}\right]^{1/(\beta+1)} \left(\frac{\nu}{xU_\infty}\right)^{\beta/(\beta+1)} \tag{20}$$

Trinh (2009) p.102 gives the values of $\alpha$ and $\beta$ as functions of $\text{Re}_x$. For pipe flow, the boundary layer thickness $\delta^+$ can be replaced by the normalised radius $R^+$.

$$R^+ = \frac{Ru_*}{\nu} = \frac{RV}{\nu}\sqrt{\frac{f}{2}} = \frac{\text{Re}\sqrt{f}}{2\sqrt{2}} \tag{21}$$

Equation (20) gives a way of relating the pipe Reynolds number to the boundary layer Reynolds number.

For $\text{Re}_x > 5000$ the data from both pipe and boundary layer flow agree well. There are no measurements of boundary layer thickness for $\text{Re}_x < 5000$ and that part of the graph is based solely on pipe flow data.

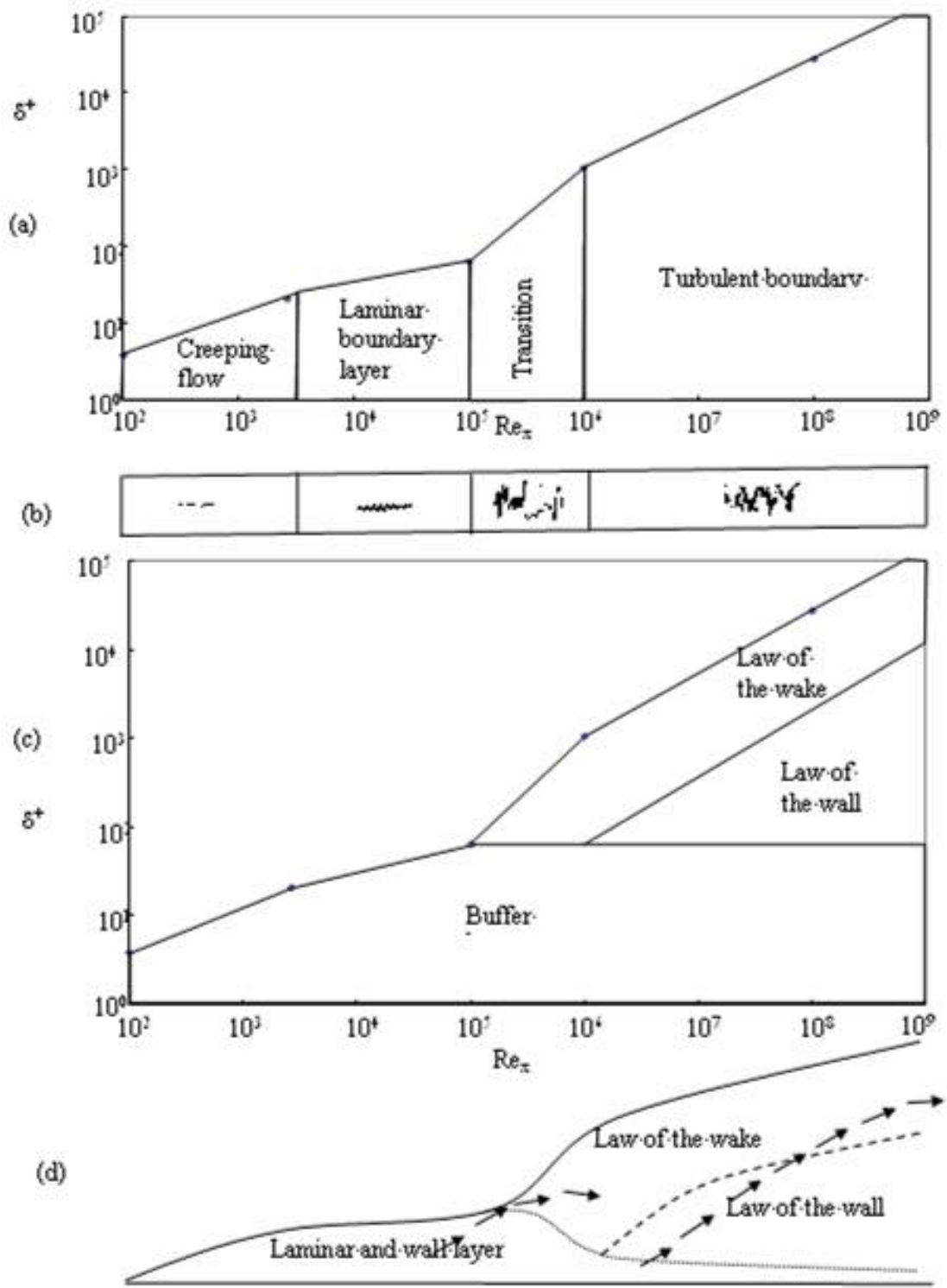

Figure 5 Representations of flow regimes (a) Reynolds flow regimes, (b) Representative velocity traces (after Schubauer and Skramstadt (1947), (c) Additive layers in normalised dimensions, (d) Physical thickness of layers (not to scale) and path of ejections in transition and fully turbulent flow.

# 4 Definition of flow regimes by additive layers

The layer described by the solution of order $\varepsilon^0$ grows with Reynolds number during laminar flow until it reaches a value $\delta_v^+ = 64.8$ for the wall layer and $\delta_b^+ = 30$ for the buffer layer at the critical Reynolds number $\text{Re}_x = 10^5$. At that point the streaming flow penetrates the quasi inviscid outer region and the law of the wake makes its appearance. The log-law layer appears at $\text{Re}_x = 10^6$. While the log-law and law of the wake layers grow with $\text{Re}_x$ the thickness of the layer $\varepsilon^0$ remained constant after $\text{Re}_x = 10^5$. Thus in this visualisation the transition region is found in the range $\text{Re}_x \approx 10^5 - 10^6$ where the velocity profile is composed of only two layers: the solution of order $\varepsilon^0$ and the law of the wake. Full turbulence sets in when the log-law starts to grow.

# 5 Discussion

We note that the boundary between the wall layer and the log-law can also be obtained by matching equation (12) with the experimental velocity profile (Figure 4). The results are identical to those obtained by the phenomenological method described in section 3. Thus the solution of order $\varepsilon^0$, equation (11) fits the time-averaged velocity in the wall layer (Trinh, 2009, p. 84) and gives a good predictions of most of the statistics on the wall layer (Trinh, 2009, p.48-80). Prandtl (1935) argued that the log-law can be derived by postulating that the typical time scale in turbulent flow called the mixing-length is proportional to the normal distance from the wall

$$l = \kappa y \tag{22}$$

where $\kappa$ is called Karman's constant. The velocity gradient becomes

$$\frac{dU^+}{dy^+} = \frac{1}{\kappa y^+} \tag{23}$$

Doshi and Gill (1970) and Trinh (1992) show that Prandtl's postulates begins with expanding the local velocity in a Taylor series

$$\Delta u = l \frac{\partial u}{\partial y} + \frac{l^2}{2!} \frac{\partial^2 u}{\partial y^2} + \frac{l^3}{3!} \frac{\partial^3 u}{\partial y^3} \ldots \quad (24)$$

and neglecting terms of 2$^{nd}$ and higher order. Therefore the log-law ceases to apply when the terms of higher order can no longer be neglected. A crude estimate of the outer limit of applicability of the log-law is obtained by setting

$$l^+ \frac{dU^+}{dy^+} = \frac{{l^+}^2}{2} \frac{d^2 U^+}{{dy^+}^2} \quad (25)$$

giving (Doshi & Gill, 1970; Trinh, 1992)

$$l^+ = 2 \frac{dU^+/dy^+}{d^2 U^+/{dy^+}^2} \quad (26)$$

which can be recognised as Karman's similarity law (1934). Equation (26) gives an alternate method for determining the boundary between the log-law and the law-of-the wake layers (Trinh, 1992).

We should note that the solution of order $\varepsilon^0$ described the subset of the NS equations (10) is independent of the solution of order $\varepsilon$. This layer, called Stokes flow is often found embedded into many types of more complex flow but it can also describe purely laminar flow in the absence of a streaming flow. The solution of order $\varepsilon^0$ ceases to apply at the point of bursting which occurs at a length L of the low-speed streak such that the kinetic energy of the streaming flow is sufficient to eject the fluid in the low speed streaks from the wall. This corresponds to a critical value of $\varepsilon$ which can be expressed as

$$\varepsilon = \frac{U_e}{L\omega} = \left(\frac{U_e L}{\nu}\right)\left(\frac{\nu}{L^2 \omega}\right) \quad (27)$$

Equation (27) indicates that the factor $\varepsilon$ is a better criterion for the onset of the ejections i.e. turbulence than the Reynolds number alone. In fact it has been shown experimentally that turbulence can be triggered at lower Reynolds numbers by introducing disturbances through the use of a trip wire (Schlichting 1960, p. 39). On the other hand Ekman (1910), for example, has shown that laminar pipe flow can be sustained up to a Reynolds number of 40,000 by carefully suppressing disturbance at the pipe entrance. Similarly Draad, Kuiken & Nieuwstadt (1998) can preserve laminar flow up to Re = 60,000 but above 14,300 Coriolis forces distort the velocity profile.

More interestingly, Draad et al. have introduced periodic disturbances of different amplitudes and frequencies to trigger turbulence in previously laminar flows. The critical amplitudes and frequencies were a function of the Reynolds number of the main flow. In such the parameter $\varepsilon$ is clearly a better indicator of transition than the Reynolds number alone.

The present visualisation shares the classical belief that the Reynolds stresses are the distinguishing feature of turbulence but further argues that there are two types of Reynolds stresses and only the fast Reynolds stresses identified with the widely observed ejections of wall fluid are characteristic of turbulence. The direct consequence of these ejections is to disturb the hitherto quasi-inviscid region beyond the solution of order $\varepsilon^0$, the wall region and thus dramatically increase in boundary layer thickness. Thus the addition of new layers to the wall layer can be used to accurate define the flow regimes. By monitoring the gradual appearance of the law of the wake and log-law layers it is possible to define clearly the extend of the transition regime. Analysis of literature data indicates that the critical Reynolds number is very dependent on flow geometry because of the choice of the characteristic dimension $L$ and the characteristic velocity (average discharge flow rate for pipes and parallel plates, approach for external flows, phase velocity for oscillating flow, Trinh 1992) but the normalised critical wall layer thickness $\delta_v^+$ is quite insensitive to flow geometry and gives a more convenient assessment of the start of transition.

## 6 Conclusion

The ejections of wall fluid in the bursting process are taken as the defining feature of turbulence. They result in the addition of two other layers to the flow field when turbulence sets in. The emergence of these layers gives a convenient way of classifying flow regimes.